# A Quench Detection and Monitoring System for Superconducting Magnets at Fermilab

A. Galt, O. Al Atassi, G. Chlachidze, T. Cummings, S. Feher, A. Hocker, S. Kotelnikov, M. Lamm, A. Makulski, J. Nogiec, D. Orris, R. Pilipenko, M. Tartaglia

*Abstract*— A quench detection system was developed for protecting and monitoring the superconducting solenoids for the Muon-to-Electron Conversion Experiment (Mu2e) at Fermilab. The quench system was designed for a high level of dependability and long-term continuous operation. It is based on three tiers: Tier-I, FPGA-based Digital Quench Detection (DQD); Tier-II, Analog Quench Detection (AQD); and Tier-3, the quench controls and data management system. The Tier-I and Tier-II are completely independent and fully redundant systems. The Tier-3 system is based on National Instruments (NI) C-RIO and provides the user interface for quench controls and data management. It is independent from Tiers I & II. The DQD provides both quench detection and quench characterization (monitoring) capability. Both DQD and AQD have built-in high voltage isolation and user programmable gains and attenuations. The DQD and AQD also includes user configured current dependent thresholding and validation times.

A 1st article of the three-tier system was fully implemented on the new Fermilab magnet test stand for the HL-LHC Accelerator Upgrade Project (AUP). It successfully provided quench protection and monitoring (QPM) for a cold superconducting bus test in November 2020. The Mu2e quench detection design has since been implemented for production testing of the AUP magnets. A detailed description of the system along with results from the AUP superconducting bus test will be presented.

## I. INTRODUCTION

The Muon-to-Electron Conversion Experiment (Mu2e) is a particle physics experiment at Fermilab designed to search for the rare process of direct muon to electron conversion in the field of a nucleus. Three large superconducting warm-bore solenoids makeup the magnet system required for this experiment: The Production Solenoid (PS); the Transport Solenoid (TS); and the Detector Solenoid (DS) [1]. The total stored energy at operating current is ~107 MJ.

A highly reliable quench protection system is required to protect the Mu2e superconducting magnets, leads, and bus. The primary requirement for the design of this system is to ensure that a true quench of any superconducting component will always be detected. It must also be designed to minimize downtime and any unnecessary interruptions to the operation of the experiment, which requires 24/7 availability to acquire the necessary data. `The quench detection system monitors the superconducting magnet coils and bus, the superconducting bus in the transfer lines, the high temperature superconducting (HTS) power and trim leads, and the vapor-cooled copper power main and trim leads. Figure 1 shows a model of the Mu2e solenoid system. The four Mu2e magnets are shown along with their respective transfer lines and feedboxes. The power leads are located on top of the feedboxes. All voltage taps are brought out through the instrumentation trees located near the feedboxes. The quench detection system resides in the same area.

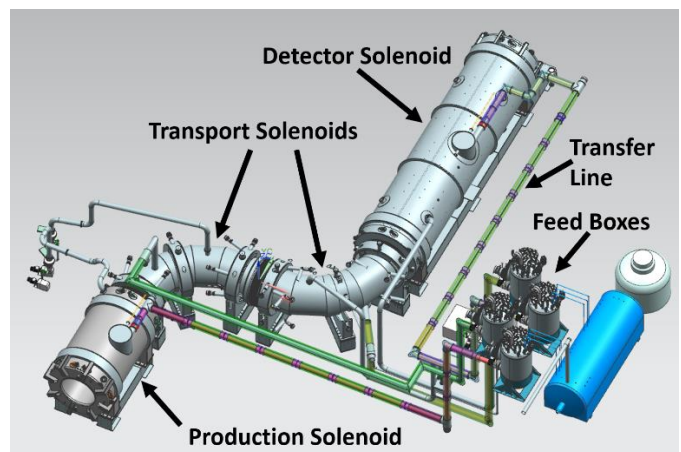

*Figure 1: Mu2e Solenoids System*

The following design criteria was incorporated in the Mu2e quench detection system to meet the requirements:
- Use of a heterogeneous quench detection design by implementing both an FPGA-based Primary (DQD) and an analog circuit design for the Redundant (AQD).
- Complete redundancy of the quench protection system from the physical voltage tap through to the power system and energy extraction (dump) system.
- Minimal connections and interfaces. Each quench detector channel has built-in isolation and signal conditioning.
- Failsafe design – open circuit protection, etc.
- Quench detection PCBs are assembled following the IPC Class 3 assembly process for high reliability electronics.
- The hardware quench detectors do not depend on any software system during operation.
- The Tier-3 is software-based (National Instruments C-RIO) but can be rebooted, if necessary, without triggering the quench detection system.
- The quench system is powered by clean power with UPS. Automatic transfer switches (ATS) switch power on UPS failure to the good UPS.

## II. QUENCH PROTECTION SYSTEM DESIGN

### A. System Block Diagram

The block diagram of the quench protection system is shown in Figure 2. Independent voltage tap (VT) cables are wired to the instrumentation tree. One set of primary DQD VT, is connected to the Digital Quench Detection while another set of redundant AQD VT, is connected to the Analog Quench Detection for complete redundancy.

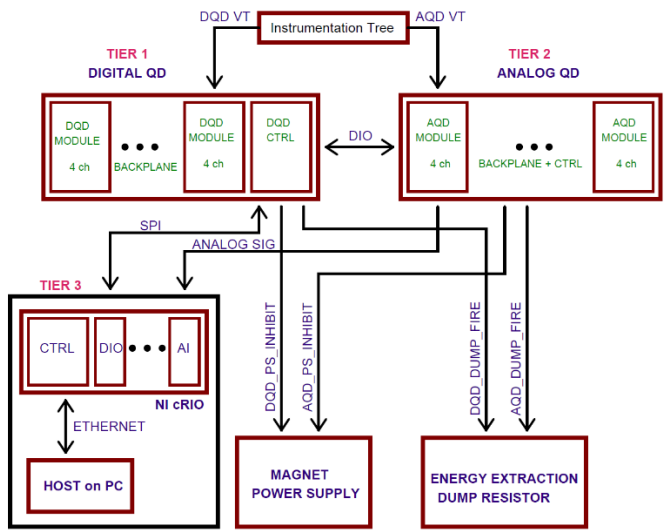

Figure 2: Quench Protection System Block Diagram

Both systems have high voltage isolation at their inputs since some VT pairs can generate up to 600 V during energy extraction.

The quench detection design provides a very flexible platform: It can be used for detecting quenches in superconducting magnets (individual and bucked signals), superconducting buses, HTS leads, and resistive voltage threshold crossing of the vapor cooled copper power leads.

### B. Digital Quench Detection Design (Tier-1)

The Digital Quench Detection system has a modular architecture. A DQD chassis, shown in Figure 3, consists of up to 8 DQD Modules, 1 DQD Controller, and a DQD Backplane. The system can be extended up to 32 individual quench channels. Quench signals originating from the voltage taps are connected to the front panel of each DQD Module. Each module can receive up to 4 individual quench signal pairs. The backplane provides communication between all modules and the chassis controller. The controller aggregates the quench signals coming from all modules configured as quench detectors. When a quench is detected, the corresponding quench signal gets transferred from the module to the controller via the backplane. A "DQD_DUMP_FIRE" signal activates the energy extraction system and a "DQD_PS_INHIBIT" switches the superconducting magnet power supply into bypass mode.

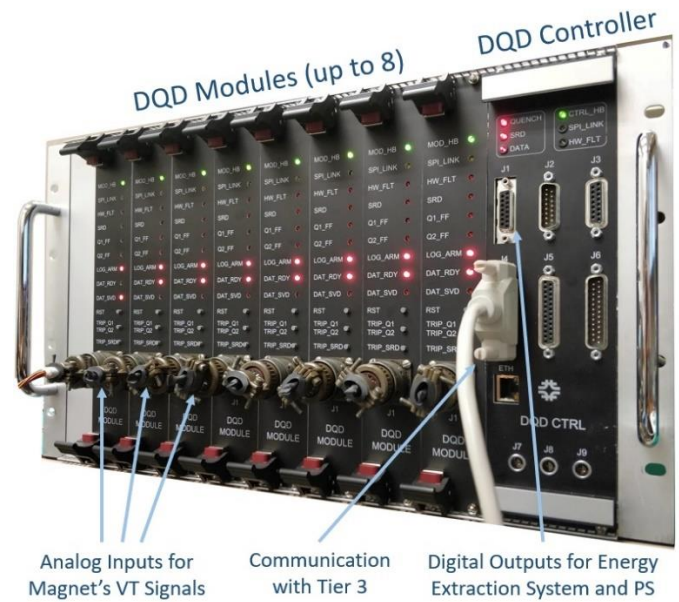

Figure 3: DQD Chassis

The block-scheme (Figure 4) illustrates the principle of the DQD Module operation. Each module has 4 individual analog channels and 4 digitally bucked channels that can buck individual channels against each other or against the current derivative (Idot). The FPGA has configurable registers that can be uploaded from the configuration file via the Tier-3 GUI. Each channel can be configured as a "QUENCH1", "QUENCH2", "SRD" (slow ramp down), or "NO_ACTION" type. If "NO_ACTION" is chosen, then this channel is considered as a quench characterization (logging) channel.

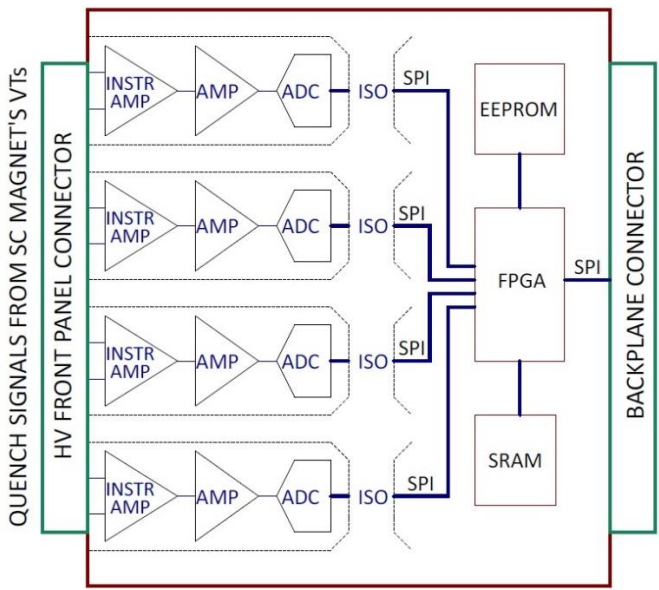

*Figure 4: DQD Module Block-Scheme*

All individual and bucked channels are digitized with a 10 kHz sample rate and saved in the circular buffer, which is implemented using the external SRAM memory. The circular buffer size is 60 kS/channel (480 kS/module).

The DQD module inputs have 2kV channel-to-channel and 2kV channel-to-ground galvanic isolation for safety and noise rejection. A programmable voltage divider (not shown) is used to attenuate input signals. The signals are then applied to an antialiasing RC filter front end of the instrumentation amplifier. A digitally configurable amplifier is used for conditioning low voltage signals; such as the signals from the low temperature superconducting (LTS) bus or the high temperature superconducting (HTS) power leads. Typical thresholds for detecting quenches in a superconducting bus is 10 mV, and 1 mV for the HTS leads. After additional conditioning, the signal is then digitized by the 16-bit ADC.

The digitized signal is sent to FPGA via the digitally isolated SPI interface where it is compared with the current dependable quench threshold specified in the "CUR_DEP_THRESH" register. If the signal exceeds the threshold longer than specified in the "VALID_TIME" register, then a quench signal is generated with the delay specified in the "DELAY_TIME" register.

### C. Analog Quench Detection Design (Tier-2)

The Analog Quench Detector (AQD) is built with a proven analog component architecture. The system works in parallel with the DQD to provide quench protection redundancy. The AQD has both channel-to-channel and channel-to-ground of 2kV galvanic isolation for safety and noise rejection. To accommodate a wide range of input voltages, each channel has jumper programmable gains ranging from 0.25 to 256. Each channel also has a jumper programmable validation time to reject false quench signals, such as noise spikes. Neighboring channels can be configured to buck coil signals and can be trimmed to compensate for inductance variation between coil segments. The AQD features hardware fault detection that monitors power supply voltages, temperature, and input overvoltage.

Trip threshold and balance are set via potentiometers on the front panel for each channel. The front panel also displays the trip and hardware fault status of each channel, as well as the configured gains setting. Each AQD chassis can accommodate seven, four-channel modules. The trip and fault status signals of each module is relayed via 10kHz modulated signal to the backplane CPLD. The backplane CPLD consolidates these signals and drives modulated output signals as configured in the firmware.

### D. Automated QC Testbench

An automated testbench was developed to thoroughly check the functionality of both the AQD and DQD modules as received from vendors, post assembly. The NI cRIO based system can exhaustively test configuration parameters for both types of modules. The system tests gains, trip latency, validation time, hardware faults, input offset, input noise, and bucking functionality.

The testbench for the DQD uses an SPI interface to configure FPGA registers, while the AQD uses isolated solid-state relays to toggle jumper settings. The testbench UI generates a simple pass/fail report for the operator and a detailed log file for the engineers to review.

### E. Tier-3 Design

The Tier-3 system is based on NI cRIO and uses NI-9401 digital modules for communicating with the DQD Chassis. The purpose of the Tier-3 is to provide a GUI interface for interacting with the DQD system and automatic quench data saving. During start-up the user is prompted to select a DQD configuration file for the DQD. After the file has been successfully uploaded, the quench signal display, shown in Figure 5, pops up with 3 analog signal charts, quench protection system digital statuses, and quench signals. When a quench event occurs, the Tier-3 waits until the DQD circular buffers stop collecting data and then writes the quench data to a file that can be used later for quench data analysis.

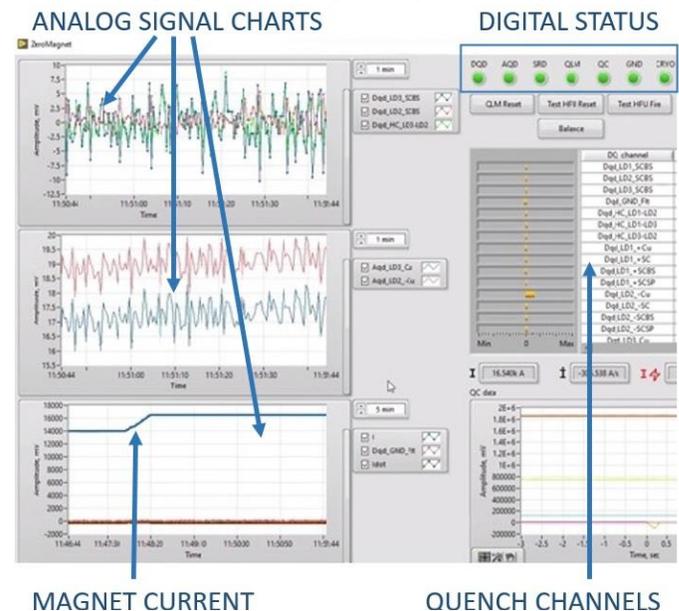

*Figure 5: Tier-3 GUI*

### III. Integration and Testing

#### A. 1st Article Implementation and Quench Testing

A Mu2e 1st Article quench detection system was implemented in a new superconducting magnet test stand under construction at Fermilab. The stand is in support of the US contribution to the High-Luminosity LHC Accelerator Upgrade Project (AUP) at CERN [2].

The test stand and quench detection system were ready for operation by November 2020. The QPM system was reviewed and an Operational Readiness Clearance (ORC) was granted for carrying out cold powered tests.

#### B. High Current Superconducting Hi-Lumi Bus Test

The test plan required power testing each of the 3 power lead configurations: Lead 1-2, Lead 3-2, and Lead 1-3. The following tests were to be performed for each power lead configuration: 1) Ramp the current in incremental steps up to the target current of 18.5kA; and 2) Ramp to 17.5kA and hold for 0.5 hours – Endurance test.

During the Lead 1-2 ramp to 18.5 kA, a quench occurred at 14,742 A. First Fault was DQD_LD2_-SCBS. Further analysis of the quench data found that the quench was located at the SC bus to vapor-cooled lead junction – see Figures 6 and 7. It was determined that the cause was additional heating at both the SC bus to Cu lead junction, as well as the external flex bus to Cu Flag junction in the lead 2. The problem was mitigated by improving the surface contact to the external junction and by increasing the liquid helium level to remove excess heat at the internal junction. A ramp to > 17 kA as well as a successful endurance run at 16.5 kA followed. A ramp to 18.5 kA and an endurance test at 17.5 kA was successful in the leads 1-3. The operation and functionality of this system proved to be a successful test of the 1st article QPM.

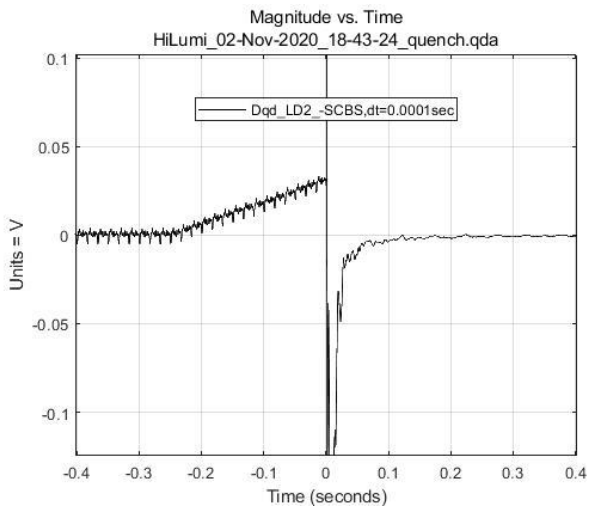

*Figure 6: Quench voltage of SC bus*

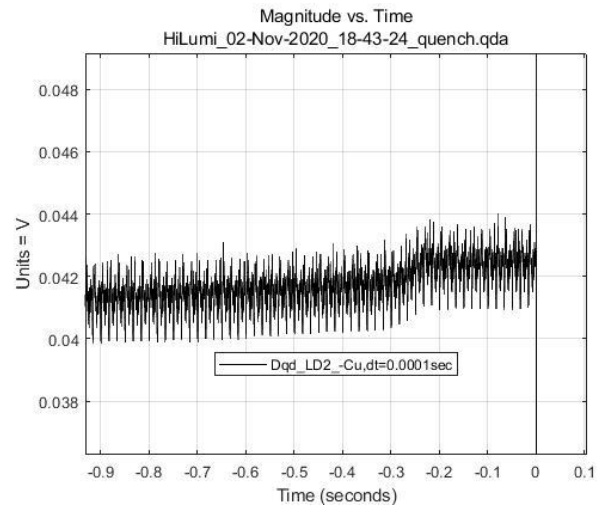

*Figure 6: Quench voltage of Cu lead including the SC bus junction.*

#### C. Past and Future Fermilab Quench Detection Applications

The design philosophy of a primary DQD and redundant AQD began with the construction of the Vertical Magnet Test Facility (VMTF) in 1997 [3]. It proved to be highly reliable; never failing to detect a quench in its >24 years of operation. Since then, DQD was implemented in FPGA using National Instruments C-RIO for other Fermilab magnet test facilities [4][5][6].

Mu2e requirements, however, drove the design for higher reliability. Work on the Mu2e QPM is now in the final production stage and will soon be integrated into the experiment.

Progress on the AUP Hi-Lumi test stand QPM was expanded to accommodate the preproduction magnet tests with additional quench detection and characterization channels.

Also, in progress, is the construction of a new High Field Vertical Magnet Test Facility (HFVMTF) at Fermilab [7], which is incorporating the Mu2e QPM design.

### IV. Conclusion

A quench protection and monitoring system was developed with special attention to fulfilling the high dependability requirements for Mu2e. This system incorporated several design features to enhance long term dependable operation. The Mu2e system is based on a 3-Tier design with complete redundancy of the quench detection system from the quench sensor through to the power supply and energy extraction systems. The Mu2e 1st article QPM was fully integrated into the AUP Hi-Lumi test stand for protecting the three superconducting LTS bus and vapor-cooled copper leads to currents >18 kA during the first cold power tests, which was successfully carried out.

The Mu2e QPM system is currently in the final construction phase and expected to be integrated into the solenoids systems at the experiment in 2022. This design is also being implemented in the design of the Fermilab HFVMTF.